\documentclass{article}
\usepackage[utf8]{inputenc}
\usepackage[a4paper, total={7in, 8in}]{geometry}
    \usepackage{graphicx}
    \usepackage{subcaption}
    \usepackage[export]{adjustbox}
    \usepackage{wrapfig}
    \usepackage{float}
    \usepackage[format=plain,labelfont={bf,it},textfont=it]{caption}
\graphicspath{ {./images/} }

\usepackage[backend=biber,style=apa]{biblatex} 
\usepackage{csquotes}
\usepackage{dirtytalk}
\usepackage{comment}

\usepackage{authblk}

\usepackage{color,soul}

\usepackage[draft]{hyperref}
\hypersetup{
    colorlinks=true,
    linkcolor=blue,
    filecolor=magenta,      
    urlcolor=black,
    citecolor=black,
}

\title{Effects of a Mixed Reality Headset on the Delay of Visually Evoked Potentials}
\author[1,*]{V\'{i}ctor Manuel Hidalgo}
\author[1,2]{Carlos Andrés Bazaes}
\author[2]{Juan-Carlos Letelier}
\affil[1]{Research and Development Department,
DTS SpA, National Aeronautical Company of Chile}
\affil[2]{Department of Biology, School of Sciences, University of Chile}
\affil[*]{Corresponding author: Victor Manuel Hidalgo, vmhidalgo@ieee.org}
\date{}

\begin{document}

\maketitle
\begin{abstract}
    Virtual and mixed reality (VR, MR) technologies offer a powerful solution for on-the-ground flight training curricula. While these technologies offer safer and cheaper instructional programs, it is still unclear how they impact neuronal brain dynamics. Indeed, MR simulations engage students in a strange mix of incongruous visual, somatosensory and vestibular sensory input. Characterizing brain dynamics during MR simulation is important for understanding cognitive processes during virtual flight training. To this end, we studied the delays introduced in the neuronal stream from the retina to the visual cortex when presented with visual stimuli using a Varjo-XR3 headset. We recorded cortical visual evoked potentials (VEPs) from 6 subjects under two conditions. First, we recorded normal VEPs triggered by short flashes. Second, we recorded VEPs triggered by an internal image of the flashes produced by the Varjo-XR3 headset. All subjects had used the headset before and were familiar with immersive experiences. Our results show mixed-reality stimulation imposes a small, but consistent, 4 [ms] processing delay in the N2-VEP component during MR stimulation as compared to direct stimulation. Also we found that VEP amplitudes during MR stimulation were also decreased. These results suggest that visual cognition during mixed-reality training is delayed, not only by the unavoidable hardware/software processing delays of the headset and the attached computer, but also by an extra biological delay induced by the headset’s limited visual display in terms of image intensity and contrast. As flight training is a demanding task, this study measures visual signal latency to better understand how MR affects the sensation of immersion.
\end{abstract}

\section{Introduction}

Mixed reality simulators are a promising solution for on-the-ground flight training given the feature that allows interaction with both virtual and real environments. Applying mixed reality is still relatively new, but using this approach for training has great potential in multiple areas and has the added value of reduced risk and cost. However, the effects of virtual reality experiences on the brain are not fully understood. There is some research on the use of EEG in Virtual Reality \parencite{EEG_VR}, but how mixed reality alters the visual experience of subjects -- thus affecting training -- remains an open question. Measuring visual evoked potentials (VEPs) is one approach that has been heavily researched to recognize the effects of diverse variables such as stimulus intensity \parencite{VEP_alterations}, anomalies in the subject's eyes \parencite{VEP_Myopia}, and stimulus adaption \parencite{VEP_dark_adaptation}. This type of measurement was thus chosen to investigate the effects of a mixed-reality Head Mounted Display (HMD) on visual processing. We found that visual evoked potentials are significantly delayed when elicited by a mixed reality HMD, suggesting that there is an extra biological delay in experience when using mixed reality systems. 

\begin{figure}[h!]
    \centering
    \includegraphics[width=0.75\linewidth]{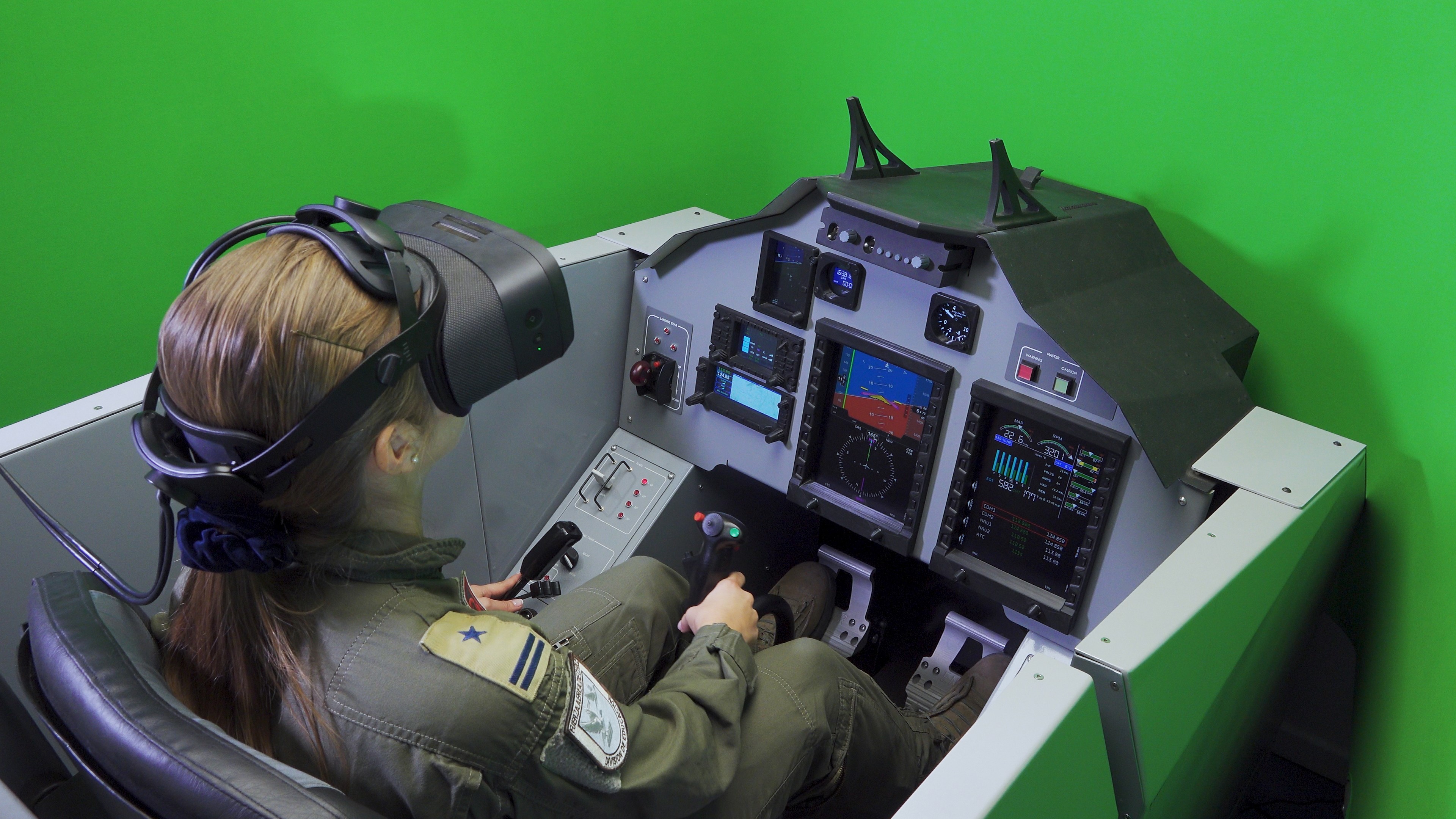}
    \caption[Flight simulator ]{\textbf{Mixed reality flight simulator MUPUN-X}. Mixed reality has the capacity to show both the real environment and a virtual scenario at the same time. In this example, the pilot can see the cockpit while a flight simulation is displayed through the headset on the green screen.}
    \label{Ejemplo_MR.png}
\end{figure}

\section{Methods}

\subsection{Subjects}
For this study, 6 subjects (5 men, 1 woman, ages=(21,22,23,23,26,64)) were recruited from the University of Chile. The experiment was approved by the "Comité de Ética de la Facultad de Ciencias de la Universidad de Chile" (Permit ID: 2225-FCS-UCH), and all subjects signed a formal consent form before any experiment was conducted.

\subsection{Experiments}
Each subject was seated at 15 [cm] in front of a light flash stimulator made by a row of 5 bright white LEDs. An Arduino Nano controlled this light source and was set to generate 100 [ms] stimuli separated by an interval uniformly distributed between 3 and 7 [s]. Visual Evoked Potentials (VEPs) were recorded using a single-channel EEG from F0-O2 under 2 conditions. In condition 1 (\emph{c1}), subjects' VEPs were recorded when they were directly stimulated without using the HMD. In condition 2 (\emph{c2}) subjects used the mixed reality headset Varjo XR-3 with the resolution set to "very-high" (1984x1984 1976x1696, 65 pixels for degree) and a 90Hz refresh rate. The hardware controlling this headset consisted of a Notebook ORIGIN EON17-X with a I9-11900K processor, a NVIDIA GeForce RTX 3080 16GB GDDR6 dedicated graphic card, and 64 GB of ram memory, under the windows 11 Pro operating system.

To record the stimulus timing we used a light sensor. In condition \emph{c1} (no HMD) the sensor directly faced the light stimulator. In condition \emph{c2} (i.e. use of the Varjo HMD) the light sensor was located inside the HMD to measure the time of appearance of the flash images in the inner uOLED Varjo displays. Subjects were instructed to look at the light stimulator for a 200 [s] time period, take a short rest, and then repeat for a new period until 160 VEPs were collected.

\begin{figure}[ht]  
    \centering
    \includegraphics[width=0.75\linewidth]{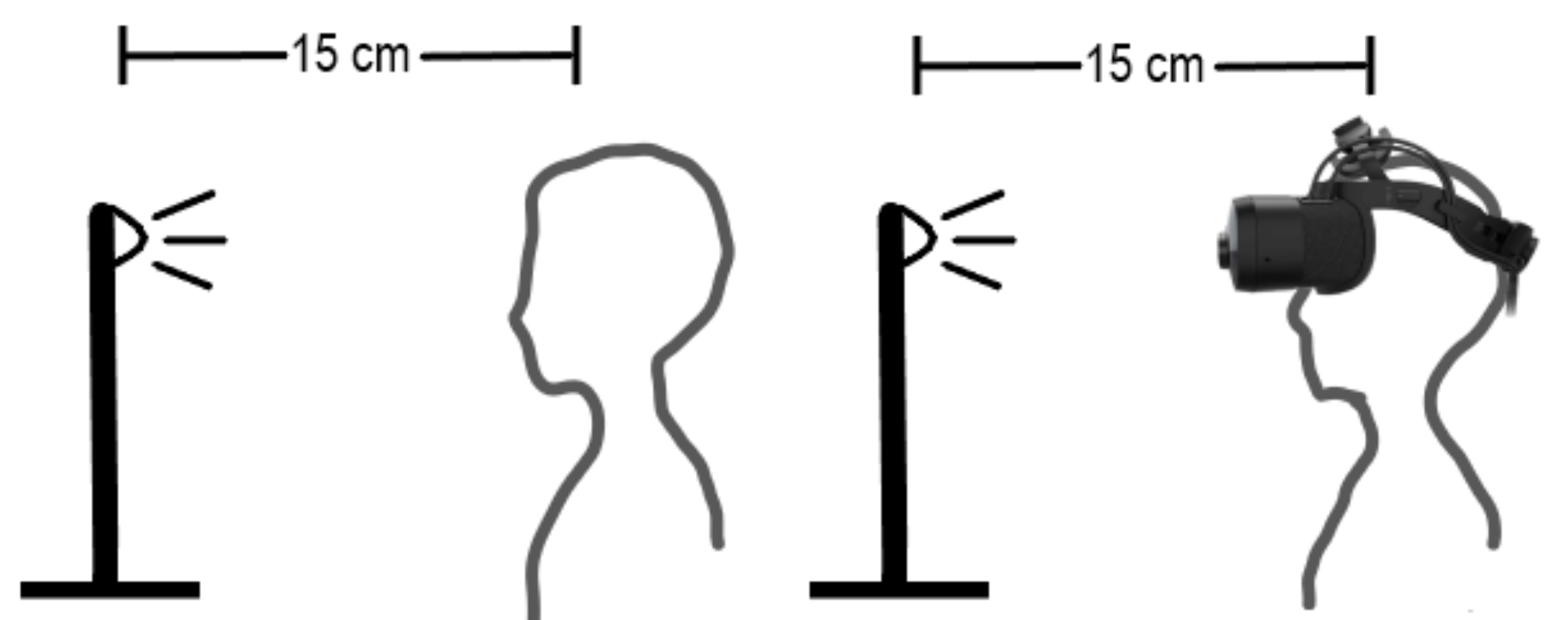}
    \caption[]{\textbf{VEP recording setup}. A  LED light source is placed 15 [cm] away from the subject and visually evoked potentials are measured in 2 conditions. Left panel (\emph{c1} ): VEPs are recorded using a direct visual stimulus. Right panel \emph{c2} : VEPs are recorded while subjects use a head mounted mixed reality display; in this case,  VEPs are elicited by the flash images on the HMD uOLED displays.}
    \label{EEG_ima.png}
\end{figure}

\subsection{Analysis}
All analyses were conducted with IGOR 9 PRO. For each VEP recording, we isolated 100 [ms] before and 300 [ms] after the stimulus from the EEG recording. Those recordings with ocular movement or with very distorted signals were set aside. From the remaining data, 75 random VEPs were selected for each subject and condition. These recordings were manually analyzed to identify the N2 peak at around 75 [ms] \parencite{VEP2004,RefVEP}. To obtain valid statistics to compare the delays of N2 peaks, we applied a resampling test (n=20) for 1000 instances for conditions and subjects \parencite{resampling}.

\section{Results}

\begin{figure}[H]
    \centering
    \includegraphics[width=1\linewidth]{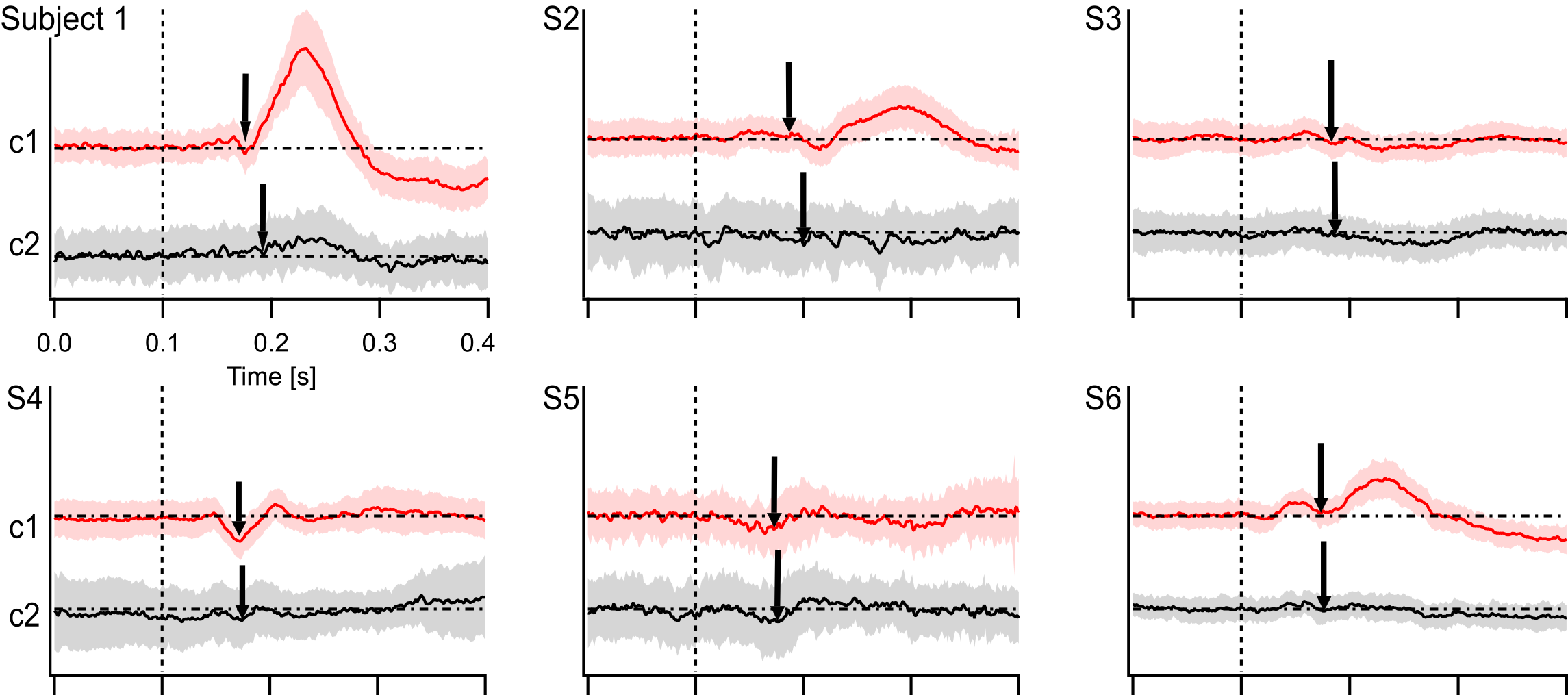}
    \caption[Visually Evoked Potencials for subject]{\textbf{Summary of Visual Evoked Potentials for each subject and each condition}. Each panel represents the collection of VEPs recorded for each subject (red=condition \emph{c1}, grey=condition \emph{c2}, heavy traces represent the ensemble average, and the diffuse colored cloud represents the standard deviation for a given time). The dotted horizontal line indicates the 0 value for the corresponding VEPs for conditions \emph{c1} and \emph{c2}. The dashed line represents stimulus onset time (\emph{c1}: LEDs ON, \emph{c2}: internal image ON). Arrows indicate the N2 peak for each subject}
    \label{EEG_resum.png}
\end{figure}

The distribution of all visual evoked potentials (VEPs) for each condition and each subject is shown in \autoref{EEG_resum.png}. In general, the distribution of condition 2 has a lower amplitude and a greater standard deviation. Furthermore delays introduced by the normal processing of the visual image (i.e. initial capture of the external image plus image processing in the host computer and the final download onto internal displays) are around 35 [ms] (data not shown). 

\begin{figure}[H]
    \centering
    \includegraphics[width=1\linewidth]{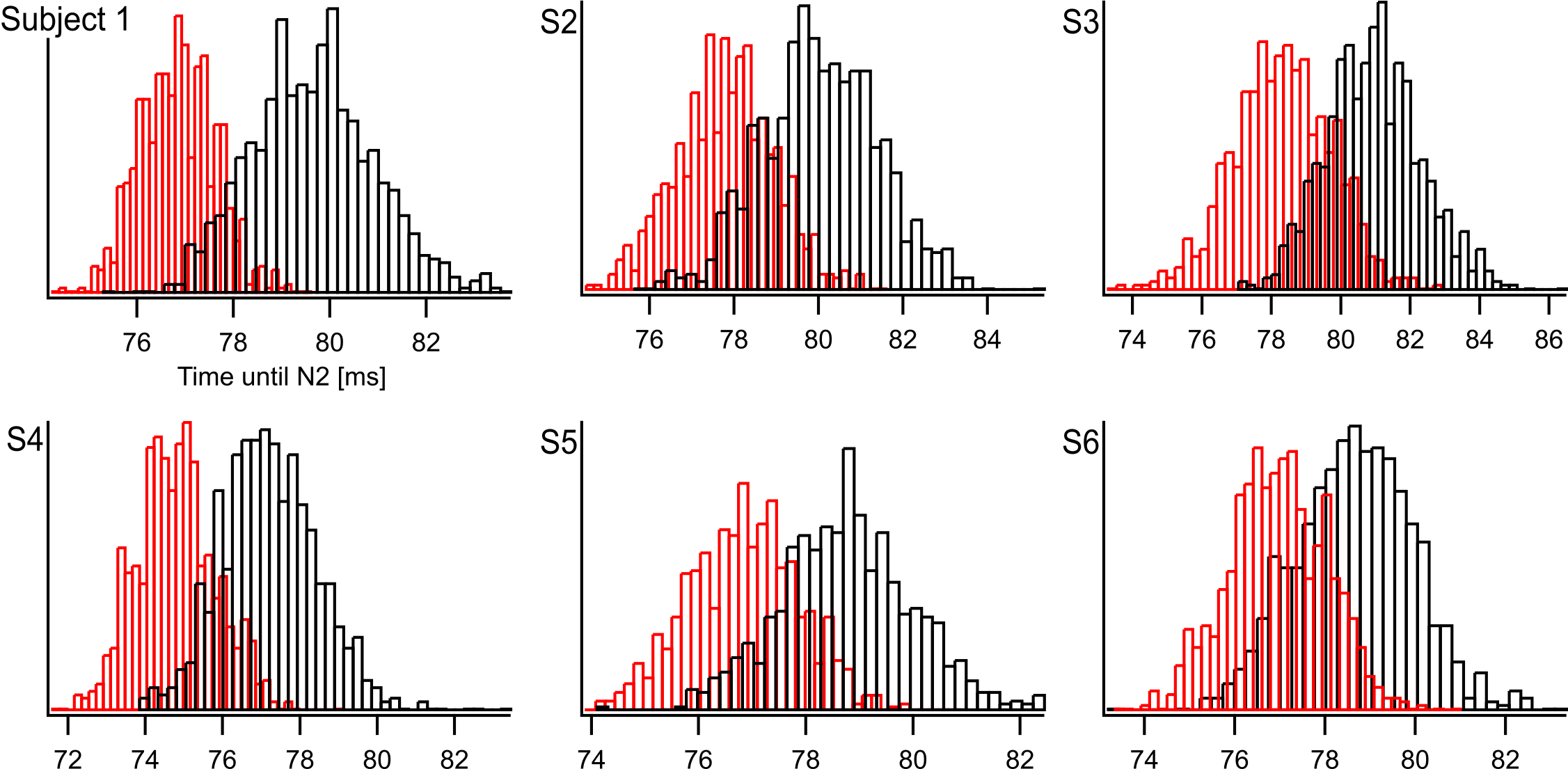}
    \caption[Resampling of temporal location of N2 peaks]{\textbf{Resampling distributions of N2 latency for each subject and experimental condition.} Each panel shows the histogram for the mean of 20 resampled N2 latency samples (1000 resampling instances) for each condition (Condition 1 is represented in red and Condition 2 in black). A two-tailed two-sample t-test at $\alpha=0.05$ with Bonferroni correction rejects the null hypothesis $\mu_{C1}=\mu_{C2}$ for all subjects.}
    \label{EEG_HIST.png}
\end{figure}

To visualize if there is a difference in the perception of the stimuli during use of the HMD, we compared the delays of the N2 peak of each VEP condition \emph{c1} and \emph{c2}, then applied a resampling test for each condition and subject (data is summarized in \autoref{EEG_HIST.png}). 
We then applied a Two-Sample t-Test with a Bonferroni correction $\alpha = 0.05 $ for each subject to compare the latency data under the two conditions and reject the null hypothesis ($H_0:$ Latencies are the same).

\section{Discussion}
In this work, we obtained EEG recordings to explore the effects of a mixed-reality HMD on the delay of visual evoked potentials (VEPs) triggered by short flashes. We showed that a high performance HMD lens introduces a small delay in visual processing. We found that the N2 peak latency was significantly delayed, by around 4 ms when the HMD was used. This difference in N2 could be related with a reduction in the intensity or in the contrast of the HMD images of the external flashes \parencite{VEP_alterations}.
Mixed reality is a promising technology for virtual flight training but its effects on visual and especially cortical processing are not fully understood. We hypothesized that the small change in latency introduced by HMD is related to the drastic change occurring with respect to the intensity and contrast of the real image vis-a-vis the image displayed. As VEPs reflect the activity of the underlying neuronal populations, it is not surprising that VEPs triggered by stimuli presented in the VARJO uOLED internal displays  have less amplitude and exhibit a slightly larger latency \parencite{VEP_alterations}. This result suggests that mixed reality training imposes a small, but consistent, delay in activating
 the visual cortex. In addition, each HMD introduces a digital processing delay due to the unavoidable steps of image capture; in the case of VARJO XR-3 this delay is around 35 [ms]. Overall, our result shows that a high performance HMD introduces delays due to hardware/software issues and neuronal dynamics. The neuronal delay albeit not negligible is small enough to enable the operational closure needed to produce the sensation of immersion. If we measure the observed delay in terms of frame rate (90Hz is equivalent to a 11.1 [ms]), it would induce a maximum delay equivalent to a single video frame. In any case, sophisticated augmented reality projects demand that these optic, visual and neuronal parameters are measured and understood in order to produce a vivid, useful immersive learning system.

\printbibliography

\end{document}